\documentclass[manuscript]{aastex}
\begin{document}
\title{Optical polarization properties of February 2010 outburst of the blazar Mrk 421}
\author{K. K. Singh}
\affil{Physics Department, University of the Free State, Bloemfontein, South Africa 9300 and 
      {Astrophysical Sciences Division, Bhabha Atomic Research Centre, Mumbai, India 400085}}
\author{P. J. Meintjes, B. van Soelen and F. A. Ramamonjisoa}
\affil{Physics Department, University of the Free State, Bloemfontein, South Africa 9300}
\and
\author{B. Vaidya}
\affil{Centre for Astronomy, Indian Institute of Technology Indore, Khandwa Road, Simrol, Indore, India 453552}
\email{kksastro@barc.gov.in}
\begin{abstract}
In this paper, we explore the behavior of optical polarization during the multi-wavelength outburst of the blazar Mrk 421 
detected in February 2010. We use optical polarization measurements in the wavelength range 500-700 nm from SPOL observations 
available between January 1, 2010 and March 31, 2010 (MJD 55197-55286) including the period of multi-wavelength flaring 
activity detected from the source around February 16-17, 2010 (MJD 55243-55244). We also use near simultaneous optical 
and radio flux measurements from SPOL in V and R bands and OVRO at 15 GHz respectively. We find that the emissions in 
the optical and radio bands do not show any significant change in the source activity unlike at X-ray and $\gamma$--ray 
energies during the outburst. The optical and radio flux measurements are found to be consistent with the long term quiescent 
state emission of the source. Moreover, the linear polarization in the wavelength range 500-700 nm decreases to a minimum value 
of 1.6$\%$ during the X-ray and $\gamma$--ray outburst which is significantly lower than the long term average value of 
$\sim$ 4.2$\%$. The angle of polarization varies between 114$^\circ$-163$^\circ$ with a preferred average value of 
$\sim$ 137$^\circ$ during this period. We estimate the degree of polarization intrinsic to the jet taking into account the host 
galaxy contamination in R band and compare this with the theoretical synchrotron polarization estimated for a power law distribution 
of relativistic electrons gyrating in an emission region filled with ordered and chaotic magnetic fields. The intrinsic linear polarization 
estimated for different epochs during the above period is found to be consistent with the theoretical synchrotron polarization produced by 
the relativistic electrons with power law spectral index $\sim$ 2.2. We find that the behavior of optical polarization possibly supports 
the two emission zone hypothesis for blazars in which X-ray and $\gamma$--rays are produced in one region whereas the optical emission 
takes place from another region permeated with ordered and chaotic magnetic fields. The decrease in linear polarization during the X-ray and 
$\gamma$--ray outburst can be attributed to the sudden dominance of chaotic magnetic field over the ordered field in the optical emission region 
in the relativistic jet of the blazar Mrk 421.
\end{abstract}

\keywords{Galaxies: active ---BL Lacertae objects: individual (Mrk 421)---optical:general---radiation mechanism: nonthermal}


\section{Introduction}
BL Lacertae objects (BL Lacs) and Flat Spectrum Radio Quasars (FSRQs) collectively form the blazar class of active galactic 
nuclei (AGN) with highly collimated relativistic jets pointing at small angles ($\le$ 10$^\circ$) towards the observer on 
Earth \citep{Urry1995}. FSRQs are observed to be more luminous and powerful than BL Lacs. In addition, the optical spectra of 
FSRQs have prominent emission lines from the thermal plasma whereas no such or weak spectral lines are observed from BL Lacs. 
Blazars are observed to emit radiation over the entire electromagnetic spectrum from radio to TeV $\gamma$--rays. 
Most of the electromagnetic radiation observed from blazars is explained by the synchrotron and inverse Compton 
processes. The broadband spectral energy distribution (SED) of blazars exhibits two broad humps peaking at low 
(radio-UV/optical-soft X-ray) and high (hard X-ray-MeV-GeV-TeV $\gamma$--ray) energies respectively. The physical 
process for the origin of low energy hump has been completely understood and is attributed to the synchrotron 
radition of relativistic electrons and positrons in the jet magnetic field. However, the origin of high energy hump is 
not very clear and is one of the important open questions in the high energy astrophysics research
until today. In the generally accepted scenario, high energy emission from blazars is attributed to the 
inverse Compton (IC) scattering of low energy photons produced both inside and outside the jet by the same population 
of relativistic electrons responsible for the origin of low energy hump. If target photons for IC scattering are the 
synchrotron photons produced inside the jet, it is referred to as synchrotron self Compton (SSC) process under the 
leptonic scenario \citep{Maraschi1992,Ghisellini1998}. On the other hand, if the seed photon field for IC scattering is 
outside the jet, the process is known as external Compton (EC) model \citep{Dermer1992,Sikora1994}. In the alternative 
hadronic scenario, models based on proton synchrotron or secondary emission from the cascade initiated by proton-$\gamma$ 
interactions have been proposed to explain the high energy hump in the blazar SED \citep{Aharonian2002,Mucke2003,Abdo2011,Fraija2015a}. 
Blazars are also observed to exhibit broadband or orphan flaring activity (dominant at X-ray and $\gamma$--ray energies) with the highest 
output flux level varying at timescales ranging from months to few minutes \citep{Cui2004,Fraija2015b,Ackermann2016,Singh2018,Singh2019}. 
The physical processes involved in the flaring activity of these sources are not completely understood. 
\par
The broadband radiation measured from the blazars is characterized as non-thermal, variable, Doppler boosted and highly polarized. 
The observations of polarized radiation at optical wavelengths from blazars support the synchrotron process in the 
ordered magnetic field in the blazar jets. FSRQs are observed to be less polarized than the radio selected BL Lacs \citep{Fan2008}. 
The observations of high degree of polarization during the flaring and quiescent states play a very important role in exploring the 
emission mechanisms operating in the jets or outflows in blazars. The measurements of the degree and angle of polarization from blazars at 
low energies help in probing the underlying particle distribution along with the strength and configuration of magnetic field in 
the emission region of the blazar jet. Correlated variability between polarization and flux in different energy bands provides 
information about the magnetic field geometry in the blazar jet. Therefore, variability study of the polarization is important for 
exploring the jet properties and associated magnetic field structure \citep{Visvanathan1998}. It has also been observed that the polarization 
is connected to the morphology and relativistic beaming and therefore plays an important role in the unification models for quasars 
\citep{Fan2008,Zakamska2005}. Mrk 421 is a well studied BL Lac object and has been considered as an excellent candidate 
for understanding the physical mechanisms involved in the blazar jets. In this work, we use the optical polarization measurements on 
Mrk 421 observed during the extreme outburst in the X-ray and $\gamma$--ray energy bands in February 2010 to investigate the physical 
processes involved in the flaring activity. The paper is organized as follows: in Section 2, we present an overview of the 
important results based on the multi-wavelength studies of the flare detected from Mrk 421 in February 2010. In Section 3, 
we describe the observations and data sets used in this work. Section 4 describes the results of the optical and radio 
observations of the source. In Section 5, we discuss the theoretical aspects of the synchrotron polarization and compare them 
with the results obtained during the outburst. Finally, we conclude the important findings of this study in Section 6.

\section{Overview of Mrk 421 outburst in February 2010}
Mrk 421 is a very active BL Lac type blazar at redshift z=0.031 \citep{Sbarufatti2005}. The source has been observed 
to undergo broadband flaring activities occasionally since its discovery at TeV $\gamma$--rays in 1991 \citep{Punch1992}. 
Here, we present an overview of the results from the X-ray and $\gamma$--ray observations of Mrk 421 outburst detected around 
February 16-17, 2010 (MJD 55243-55244). The X-ray outburst detected by the MAXI-GSC and \emph{Swift}-BAT (Burst Alert Telescope) 
was found to be the brightest among the previous high activity states of the source \citep{Isobe2010}. The strength of the 
jet magnetic field derived from the X-ray observations was found to be weaker than the earlier reported values for this blazar. 
In the $\gamma$--ray band, the outburst was simultaneously detected by the \emph{Fermi}-LAT (Large Area Telescope) at 
MeV-GeV energies \citep{Singh2012}. Significant spectral hardening in both X-ray and MeV-GeV $\gamma$--ray bands was 
observed irrespective of the large flux variations during the outburst \citep{Isobe2010,Singh2012}. The observed spectral variation 
during this outburst was different from the results of previous flaring episodes of Mrk 421 where a positive correlation between hardness 
and flux had been detected. At TeV $\gamma$--ray energies, the flaring activity of Mrk 421 was simultaneously detected by many ground 
based $\gamma$--ray telescopes operational during the period of broadband outburst and rapid flux variations on timescales down to 
few minutes were observed on the night of February 16-17, 2010 \citep{Tluczykont2011,Fortson2012,Shukla2012,Singh2015,Bartoli2016}.

\par
Near simultaneous broadband data on the Mrk 421 outburst observed in February 2010 have been extensively used to understand the physical 
process involved in flaring activity of the source in the literature. \cite{Zheng2014} investigated the time-dependent properties of the 
outburst using a single zone SSC model and assuming a strong magnetic turbulence for the stochastic acceleration of electrons to relativistic 
energies. They concluded that the X-ray and $\gamma$--ray emissions observed during outburst were produced via synchrotron and SSC processes 
by the population of electrons whose injection into the emission region changed with time. \cite{Singh2017} used a time dependent one zone 
SSC process to model the daily X-ray and $\gamma$--ray light curves by numerically solving the kinetic equation for the evolution of the 
population of electrons in the emission region. 
The injection of electrons into the emission region was assumed to be a time-dependent power law produced by the putative acceleration 
process. It was suggested that the X-ray and $\gamma$--ray flaring activity observed in February 2010 can be attributed to an 
efficient acceleration process associated with the sudden increase in the  electron density in the emission region. 
However, the optical polarization measurements during the outburst have not been exclusively used to probe this particular 
flaring activity of the blazar Mrk 421. Radio follow-up observations of the X-ray and $\gamma$--ray flares at 22 GHz detected 
the apparent superluminal inward motion of the jet component instead of a newly born component associated 
with the outburst \citep{Niinuma2012}. The motion of the jet component toward the core was attributed to the possible 
ejection of a new component which could not be resolved during the radio observations.

\section{Data Set}
A broadband giant flaring activity was observed from the blazar Mrk 421 in February 2010 by nearly all the ground and 
space based instruments world wide. The maximum change in the flux state was detected around February 16-17, 2010 (MJD 55243-55244)
in the X-ray and $\gamma$--ray energy bands. In this work, we have used near simultaneous optical and radio archival data 
collected between January 1, 2010 and March 31, 2010 (MJD 55197-55286) and publicly avaialable under the \emph{Fermi Multiwavelength 
Observing-Support Programs}\footnote{https://fermi.gsfc.nasa.gov/ssc/observations/multi/programs.html}. The data for three 
months including the period of X-ray and $\gamma$--ray outbursts have been selected for exploring the optical and radio 
behavior of the blazar Mrk 421 during and without the flaring activity.
\par
The spectropolarimeter (SPOL) at the Steward observatory of the Univeristy of Arizona provides ground based observational support 
to the \emph{Fermi}-Large Area Telescope (LAT) monitored blazars \citep{Smith2009}. SPOL is a high throughput and moderate 
resolution dual beam instrument with a waveplate and Wollaston prism. This instrument contributes to the measurements of the 
degree of linear polarization in the wavelength range $\lambda$= 500-700 nm and optical magnitudes in V and R bands from the 
nightly monitoring of the $\gamma$--ray bright blazars. The linear polarization is measured by using a $\lambda$/2 waveplate 
in the telescope. The optical magnitude/flux measurements are provided from the differential spectrophotometry. A Johnson band 
pass filter with effective wavelength of 540 nm is used for V band differential photometry whereas R band measurements are 
performed using Kron-Cousins band pass filter with effective wavelength of 640 nm. The optical measurements by SPOL in V and 
R bands are not corrected for the interstellar extinction and contamination from the host galaxy starlight. In this work, 
we have used the polarization data and optical (V and R bands) flux measurements from the 
SPOL observations\footnote{http://james.as.arizona.edu/~psmith/Fermi/} of the blazar Mrk 421 during the period mentioned above.  
\par
The 40 m telescope at the Owens Valley Radio Observatory (OVRO) provides radio observations of selected \emph{Fermi}-blazars at 
15 GHz \citep{Richards2011}. The optics of this telescope uses a large diameter f/0.4 parabolic reflector with an altitude-azimuth 
mount. A cooled receiver operating in the Ku band with central frequency of 15 GHz and bandwidth of 3 GHz is installed at the prime 
focus of the telescope. The radiometry with the 40 m telescope includes pointing to a nearby bright source, calibration of noise 
source and flux density measurements. In the present work, we have used near simultaneous radio flux measurements at 15 GHz on  
blazar Mrk 421 available from the OVRO archive\footnote{http://www.astro.caltech.edu/ovroblazars/data.php/} for the period of 
optical observations discussed above.  

\section{Results}
\subsection{Optical and Radio Light curves}
\begin{figure}
\epsscale{.75}
\plotone{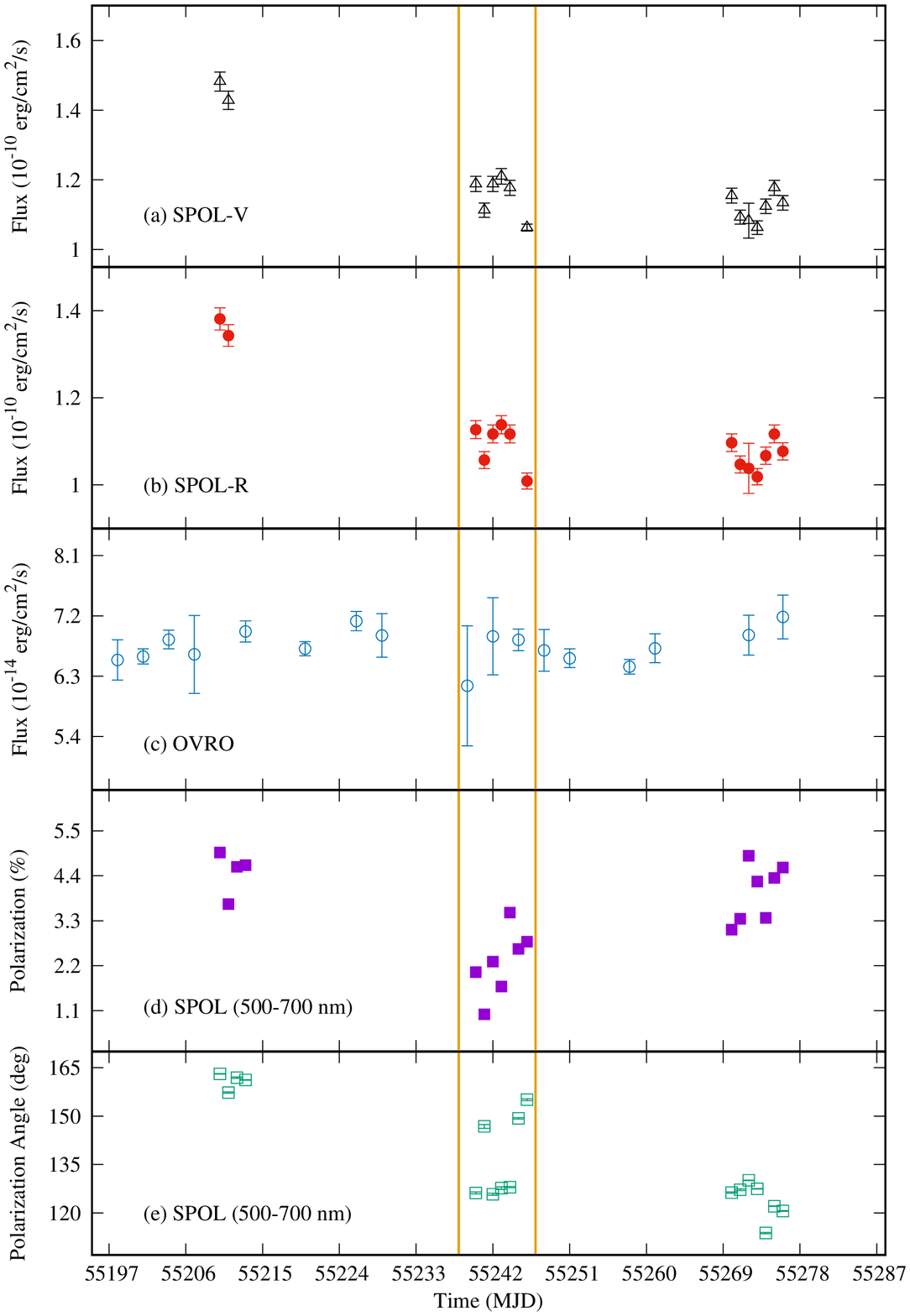}
\caption{Daily optical and radio light curves for the blazar Mrk 421 observed between January 1, 2010 and 
	 March 31, 2010 (MJD 55197-55286). The time interval between two vertical lines corresponds to the 
	 week February 13-19, 2010 (MJD 55240-55246) involving the giant X-ray and $\gamma$--ray flares observed 
	 from the source on February 16-17, 2010 (MJD 55243-55244). The error bars in the measured degree of polarization 
	 (d) and polarization angle (e) are very small.} 
\label{fig1}
\end{figure}

The daily light curves of Mrk 421 in optical and radio wave-bands for the period January 1, 2010 and March 31, 2010 
(MJD 55197-55286) are shown in Figure \ref{fig1}(a-c) respectively. The time interval between the two vertical lines 
(MJD 55240-55246) in Figure \ref{fig1} indicates the period covering the giant flaring activity of the source detected 
in the X-ray and $\gamma$--ray bands. Visual inspection of the optical light curves, Figure \ref{fig1}(a-b), suggests 
that the source was in a relatively higher emission state much before the X-ray and $\gamma$--ray outburst. During the 
period of flaring activity in high energy bands, the optical emission in both bands (V and R) is almost steady and 
consistent with the long term quiescent state of the source. It is important to note here that the optical fluxes reported in 
Figure \ref{fig1}(a-b) are not corrected for the contamination due to thermal emission from the host galaxy and interstellar 
reddening. The host galaxy of most of the blazars is observed to be relatively bright in optical bands and its contribution is 
very strong in near-infrared \citep{Nilsson2007}. The host galaxy contribution in the optical emission from Mrk 421 is discussed 
in detail in Section 5.1. The radio observations at 15 GHz from OVRO, as reported in Figure \ref{fig1}(c), also indicate a steady 
state of the source during the whole period and do not exhibit any change in the source activity during the X-ray and $\gamma$--ray 
outburst. We use $\chi^2$-test of null-hypothesis of constant emission to characterize the nature of broadband emission during the 
different epochs. During the period of outburst, the optical emission in the V and R bands is characterized by constant flux levels 
of (1.11$\pm$0.03)$\times$10$^{-10}$~erg~cm$^{-2}$~s$^{-1}$ and (1.09$\pm$0.02)$\times$10$^{-10}$~erg~cm$^{-2}$~s$^{-1}$ respectively 
using a $\chi^2$-test of null-hypothesis. Whereas the $\chi^2$-test for fluxes measured during the period excluding the flaring episode 
suggests that the optical emission in both bands is variable. This is consistent with the visual inspection of the optical light curves 
where a high optical activity state of the source is observed prior to the X-ray and $\gamma$--ray outburst. The radio emission at 15 GHz  
described by a constant flux of (6.72$\pm$0.05)$\times$10$^{-14}$~erg~cm$^{-2}$~s$^{-1}$ throughout the period considered in this study  
is compatible with the long term quiescent state of the blazar Mrk 421. Therefore, the optical and radio light curves indicate that there 
is no change in the source activity at low energies during its giant flaring activity in X-ray and $\gamma$--ray energy bands observed 
in February 2010.

\subsection{Observed Polarization}
The degree of optical linear polarization and corresponding polarization angle measured in the wavelength range 500-700 nm by the 
SPOL during the period January 1, 2010 and March 31, 2010 (MJD 55197-55286) are displayed in Figure \ref{fig1}(d-e). The polarization 
measurements between the two vertical lines are near simultaneous observations during the X-ray and $\gamma$--ray outburst of the 
blazar Mrk 421 in February 2010. It is evident from  Figure \ref{fig1}(d) that the degree of optical linear polarization decreases 
during the flaring activity in high energy bands. The maximum degree of linear polarization measured during the time interval of 
outburst is comparable to the minimum degree of polarization observed during the period without flaring episode. The degree of linear 
polarization during the peak of X-ray and $\gamma$--ray outburst observed around February 16-17, 2010 (MJD 55243-55244) is found to be 
lowest ($\sim$ 1.6$\%$). A constant fit to the degree of polarization measured during the flaring episode gives an average polarization 
of 2.23$\pm$0.31$\%$, which is only 50$\%$ of the average polarization 4.24$\pm$0.18$\%$ estimated using the measurements excluding the 
period of X-ray and $\gamma$--ray outburst. This suggests that the optical linear polarization reduces significantly during the flaring 
activity of the source. The time variation of the polarization angle shown in  Figure \ref{fig1}(e) suggests that the polarization angle 
measured during the period of this study randomly changes from $\sim$ 114$^\circ$ to a maximum of $\sim$ 163$^\circ$. During the period of 
X-ray and $\gamma$--ray outburst (time interval between the vertical lines in  Figure \ref{fig1}), the polarization angle varies between 
126$^\circ$--155$^\circ$. This indicates that the variation in polarization angle during the flaring activity is within the values measured 
for the periods before and after the outburst. The polarization angle for Mrk 421 is found to have a preferred mean value of 
$\sim$ 137$^\circ$ during January 1, 2010 and March 31, 2010 (MJD 55197-55286) and it does not change significantly during the X-ray and 
$\gamma$--ray outburst of the source. The value of the polarization angle as 180$^\circ$ indicates that the polarization vector is aligned 
along north-south direction and it increases going east of north.
\par
\begin{figure}
\epsscale{.75}
\plotone{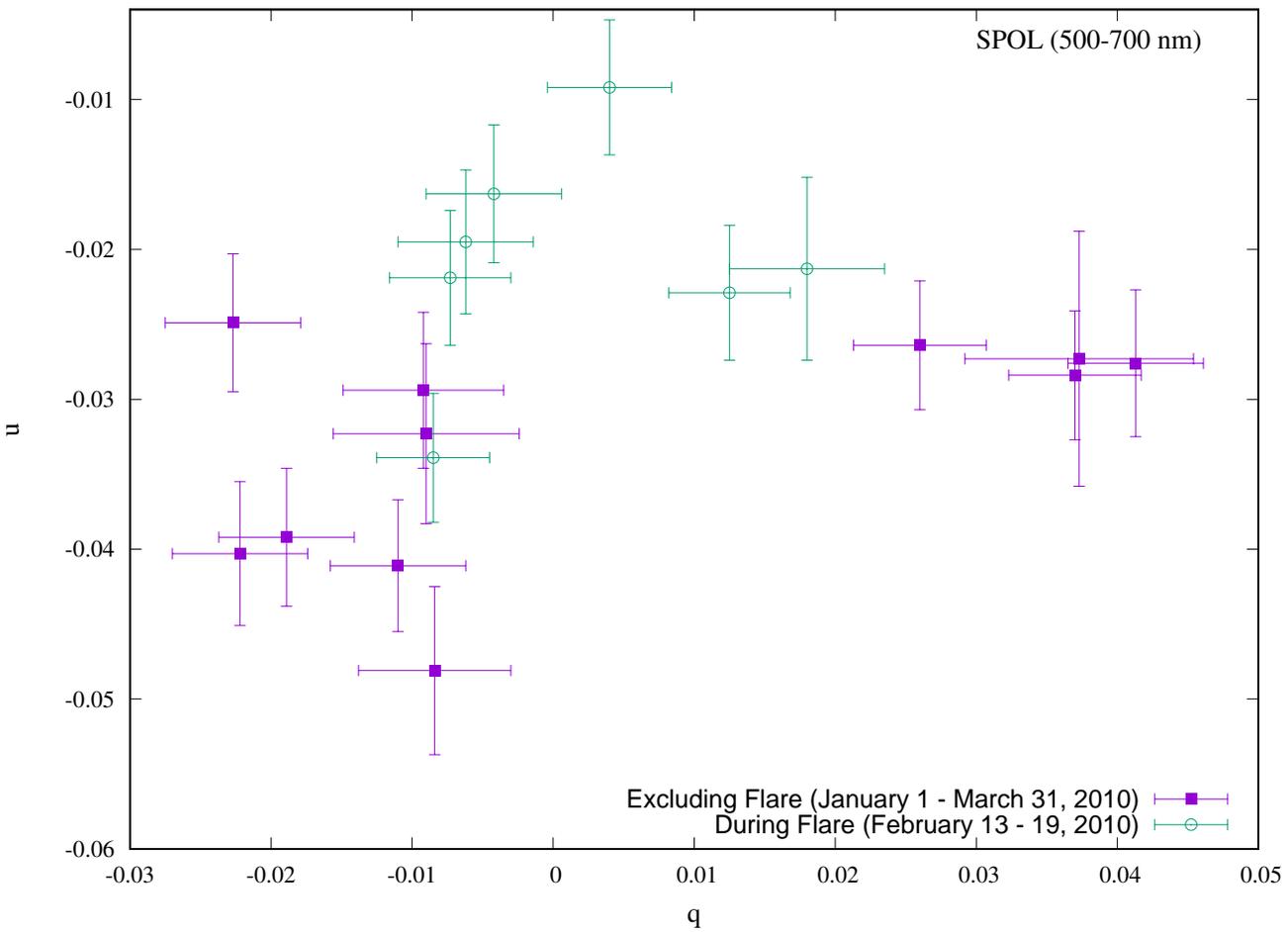}
\caption{Scatter plot for the Stokes parameter $u$ and $q$ used to estimate the linear polarization and 
	polarization angle by the SPOL in the wavelength range 500-700 nm during January 1, 2010 - March 31, 2010 (MJD 55197-55286).} 
\label{fig2}
\end{figure}

The linear polarization of incoherent synchrotron emission is complex with its real and imaginary parts as observable 
quantities. It can be expressed as \citep{Burn1966,Sokoloff1998}
\begin{equation}
	P = p~e^{i2\phi}	
\end{equation}
where $p$ and $\phi$ are the observed degree and angle of linear polarization respectively. The observable quantities 
$p$ and $\phi$ are estimated in terms of the measured Stokes parameters using the relations
\begin{equation}
	p=\sqrt{u^2+q^2}~~~~~~~\&~~~~~~ \phi=\frac{1}{2}~\arctan\left(\frac{u}{q}\right)
\end{equation}	
where $u$ and $q$ are Stokes parameters normalized by the total synchrotron intensity (I). The relation between Stokes parameters 
indicates that $u$ is a linear function of $q$ and slope of the line (tan~$2\phi$) corresponds to the angle of polarization.   
The scatter plot ($u$ versus $q$) of the measured mean values of the normalized Stokes parameters for individual observations with 
the SPOL is shown in Figure \ref{fig2} to understand the polarimetric behavior of the blazar Mrk 421 during January 1, 2010 -- March 31, 
2010 (MJD 55197-55286). The distance of individual points from the origin of $q - u$ plot gives the degree of linear polarization. 
The Stokes parameters depicted in Figure \ref{fig2} are observed to be linearly correlated and a linear fit gives the slope of the line to be 
0.13$\pm$0.10. However, no strong conclusion can be drawn from the scatter plot due to large error bars and less data points. 
The linear polarization reported in Figure \ref{fig1}(d) has been corrected for the statistical bias associated with this being a positive 
definite quantity. We note that no correction for thermal emission contamination from the host galaxy of the source is 
applied in the degree of polarization measurements by the SPOL. The effect of host galaxy contamination on the linear polarization is 
discussed in Section 5.1.  
\subsection{Variability and Correlations}
The observation of highly variable and polarized radiation at optical and radio wavelengths is one of the important characteristics of 
broadband emissions from blazars during the flaring episodes. We use three different statistical parameters to characterize 
the variability in optical and radio emissions  from the blazar Mrk 421 during the X-ray and $\gamma$--ray outbursts observed during 
February 2010 (Figure \ref{fig1}). To quantify the intrinsic variability, we first estimate the fractional variability ($F_{var}$) 
which is defined as \citep{Vaughan2003,Singh2018}
\begin{equation}
	F_{var}=\sqrt{\frac{S^2 -E^2}{F^2}}
\end{equation}
and the error in $F_{var}$ is given by
\begin{equation}
	\Delta F_{var}=\sqrt{\left(\sqrt{\frac{1}{2N}}\frac{E^2}{F^2F_{var}}\right)^2+\left(\sqrt{\frac{E^2}{N}}\frac{1}{F}\right)^2}
\end{equation}  
where $S^2$ is the variance, $E^2$ is the mean square measurement error, $F$ is the mean and $N$ is the number of measurements available 
in a given period. This parameter takes into account the uncertainties in the measurement of a physical quantity and gives degree of 
intrinsic variability of the source. The value of $F_{var}$ close to zero implies no significant variability whereas value close to one 
indicates strong variability. Next, we use the amplitude of variation ($A_{mp}$) to calculate the peak-to-peak variability, which is 
expressed as \citep{Heidt1996,Singh2018}
\begin{equation}
	A_{mp}=100\times \frac{\sqrt{(F_{max}-F_{min})^2-2\sigma^2}}{F}~~\%
\end{equation}
and the error in $A_{mp}$ is given by  
\begin{equation}
	\Delta A_{mp}=100\times \left(\frac{F_{max}-F_{min}}{FA_{mp}}\right)\sqrt{\left(\frac{\Delta F_{max}}{F}\right)^2  
	               + \left(\frac{\Delta F_{min}}{F}\right)^2 + \left(\frac{\Delta F}{F_{max}-F_{min}}\right)^2 A_{mp}^4}~~\%
\end{equation}
where $F_{max}$ and $F_{min}$ are the maximum and minimum values of the measurement with uncertainties $\Delta F_{max}$ and $\Delta F_{min}$ 
respectively, $\Delta F$ is the error in mean, and $\sigma$ is the average measurement error. The third parameter we estimate is the 
relative variability amplitude ($RVA$) or variability index, which is defined as \citep{Kovalev2005,Singh2018} 
\begin{equation}
	RVA=\frac{F_{max}-F_{min}}{F_{max}+F_{min}}    
\end{equation} 
and the uncertainty on $RVA$ is given by
\begin{equation}
	\Delta RVA=\frac{2}{(F_{max}+F_{min})^2}\sqrt{(F_{max}~\Delta F_{min})^2 + (F_{min}~\Delta F_{max})^2}
\end{equation}
The variability parameters described above have been computed for all the observables reported in Figure \ref{fig1} for two epochs namely 
during the flare and excluding the flare. The values of the above variability parameters estimated for the two epochs are given in 
Table \ref{table1} \& \ref{table2} respectively. The $F_{var}$ values calculated for optical emission (V and R bands) during the two epochs 
are nearly similar and consistent with the long term quiescent behavior of the source. This indicates that there is no sign of strong 
intrinsic variability present in the optical emissions during the X-ray and $\gamma$--ray outburst of Mrk 421. The values of $F_{var}$ for 
radio emission at 15 GHz also do not shown any significant variability. However, the values of $A_{mp}$ and $RVA$ parameters estimated for 
the optical emissions (V and R bands) during the period without flaring episode are larger than the values during outburst. This can be 
attributed to the relatively high optical activity state of the source observed prior to the X-ray and $\gamma$--ray outburst. 
The $F_{var}$ values for the degree of polarization hint that the optical polarization is more variable during the X-ray and $\gamma$--ray 
outburst than the period excluding the flaring episode. $A_{mp}$ and $RVA$ parameters which mainly depend on the peak values of the observables 
are also found to be relatively higher during the flaring episode for the linear polarization. This is consistent with the fact that the linear 
polarization is observed to decrease to its lowest value of $\sim$ 1.6$\%$ during the period of outburst. The three variability parameters 
estimated for the polarization angle during both epochs do not show any significant change in their values. The variability analysis using 
the above three parameters suggests that the optical and radio fluxes do not exhibit variability during the flaring episode. Whereas, X-ray and 
$\gamma$--ray emissions from the blazar Mrk 421 are characterized by significant variability during the period of outburst in 
February 2010 \citep{Singh2015}. This indicates that the emissions from Mrk 421 in low and high energy bands are not co-spatial during the 
giant flaring activity of the source.
\par
\begin{figure}
\epsscale{.75}
\plotone{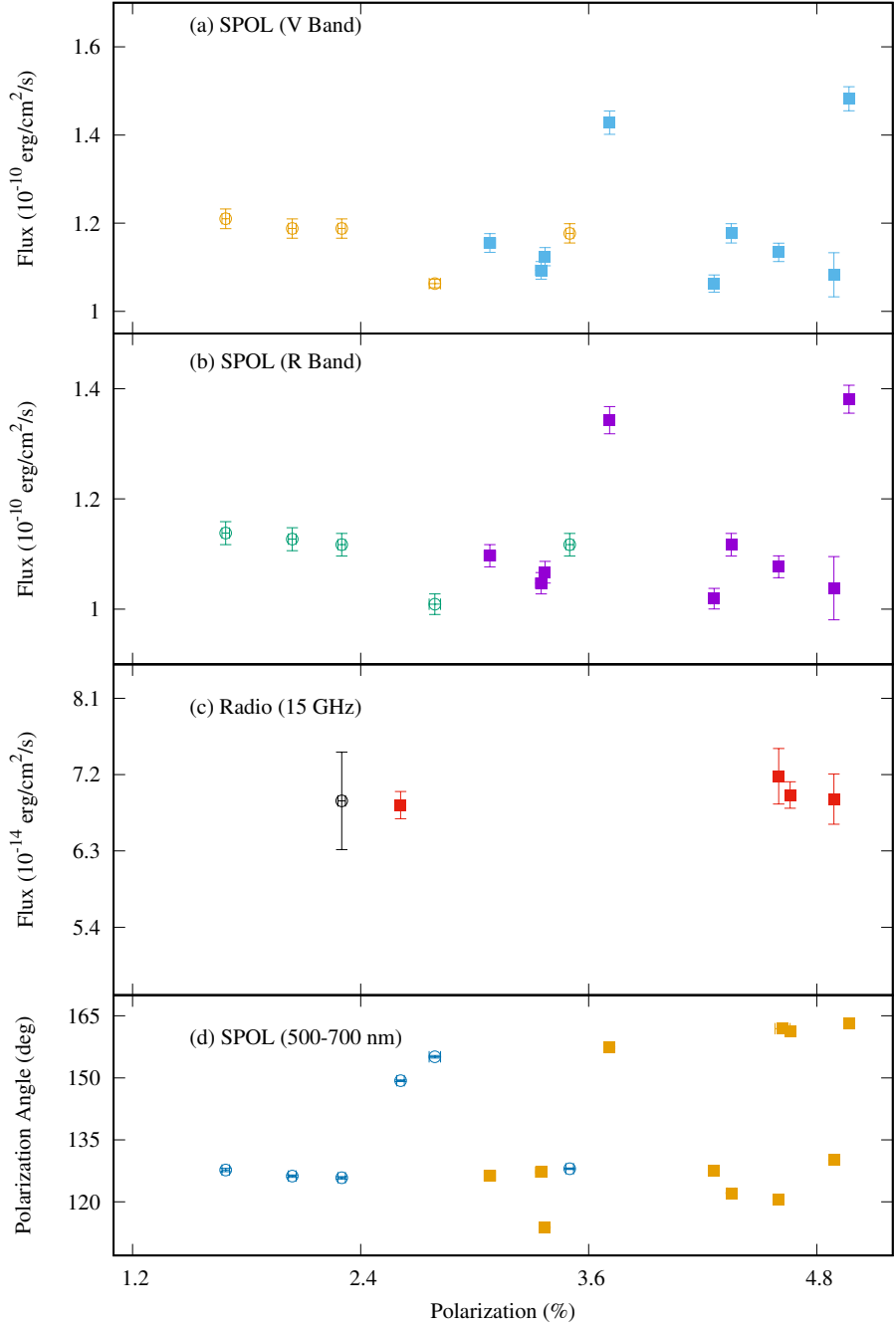}
\caption{Scatter plot for the correlations between polarization measured by SPOL telescope in the wavelength range 500-700 nm 
	and other observables measured quasi-simultaneously. The points with filled square correspond to the measurements 
	available from January 1, 2010 to March 31, 2010 (MJD 55197-55286) excluding the period of outburst from the source. 
	Observations available during the week February 13-19, 2010 (MJD 55240-55246) involving the X-ray and $\gamma$--ray 
	flare of the source are represented by the open circles.} 
\label{fig3}
\end{figure}

We also estimate the \emph{Pearson's correlations coefficients} to probe the intrinsic correlation of the optical-radio flux and polarization 
angle with the simultaneously observed degree of linear polarization. The Pearson's coefficient characterizes the linear correlation between two 
observables related to the broadband emission from the sources like blazars. The optical flux in two bands (V and R), radio flux at 15 GHz and 
angle of polarization are plotted against the near simultaneous measurement of degree of linear polarization in Figure \ref{fig3} for the 
periods during the flare and excluding the flare of the blazar Mrk 421. The visual inspection of Figure \ref{fig3} and estimated values of 
Pearson's coefficients ($\sim$0.20) both suggest that the degree of linear polarization is very weakly correlated with other observables 
during the two epochs. This is found to be compatible with the variability analysis described above for different physical quantities measured 
from the blazar Mrk 421. Very weak or no correlation has been reported between the TeV $\gamma$--ray emissions and near simultaneous optical 
and radio flux measurements during the flaring activity of Mrk 421 detected in February 2010 \citep{Singh2015}. However, a strong correlation 
between $\gamma$--ray and X-ray fluxes derived during the same period supports one zone SSC process for high energy emission from the blazar 
Mrk 421. This also suggests that the broadband emissions in low (optical and radio) and high (X-ray and $\gamma$--ray) energy bands can be 
produced from different regions in the jet of Mrk 421 during the outburst. 

\section{Discussion}

\subsection{Intrinsic Jet Polarization}
The degree of polarization is quantified as the fraction of polarized flux measured from a source. The synchrotron emissions from the 
relativistic jet of blazars are characterized by strong linear polarization in optical and radio bands. However, several depolarization 
effects like host galaxy contamination and  wavelength dependent Faraday rotation play an important role in decreasing the degree of 
synchrotron polarization. The degree of linear synchrotron polarization intrinsic to the jet is given by \citep{Carnerero2017}
\begin{equation}\label{int-pol}
	\Pi_{jet}=\frac{\Pi_{obs}}{\left(1-\frac{F_{host}}{F_{obs}}\right)}
\end{equation}	
where $\Pi_{obs}$ and $F_{obs}$ are the measured degree of linear polarization and flux in a given wavelength band respectively 
and $F_{host}$ is the unpolarized flux contribution from the host galaxy. The polarization measurement by the SPOL in the wavelength 
range 500-700 nm has an effective wavelength close to the R band. The host galaxy contribution due to the thermal emission in R band 
is estimated to be $\sim$ 3.68$\times$10$^{-11}$ erg~cm$^{-2}$~s$^{-1}$ \citep{Carnerero2017}. The average optical flux in R band observed 
from the source during the period of X-ray and $\gamma$--ray outburst is $\sim$ 1.09$\times$10$^{-10}$ erg~cm$^{-2}$~s$^{-1}$ (Section 4.1). 
This indicates that the host galaxy contribution is $\sim$ 30$\%$ of the observed flux from the blazar Mrk 421 and hence its effect on the 
depolarization of the synchrotron radiation from the jet will be significant. The intrinsic polarization to the jet in R band is found 
to be $\Pi_{jet} \approx 1.5\times ~\Pi_{obs}$ (Equation \ref{int-pol}). In V band, the contamination of host galaxy 
($F_{host} \sim$ 1.37$\times$10$^{-11}$ erg~cm$^{-2}$~s$^{-1}$) is relatively small as compared to R band \citep{Nilsson2007}. 
This implies that the host galaxy has significant depolarization effect on the linear synchrotron polarization measured from the jet of 
Mrk 421 in optical R band. In 1966, Burn proposed that the degree of polarization can be expressed as \citep{Burn1966}
\begin{equation}
	\Pi_{linear} \propto ~ e^{-\lambda^4}
\end{equation}	
where $\lambda$ is the observational wavelength. Spectropolarimetric study of blazars also suggests the wavelength dependence of the degree of 
linear polarization \citep{Kul1987,Blinov2016}. Therefore, Faraday dispersion can depolarize the synchrotron polarization due to the wavelength 
dependence of its rotation measure (RM $\propto \lambda^{-2}$) in the optical band. However, we are not able to explore the depolarization effect 
of Faraday rotation in this work due to the unavailability of the near simultaneous observations of polarization in different wavelength bands. 
According to \cite{Jones1985}, the synchrotron radiation from blazars can be depolarized due to the superposition of emissions from more than 
one region in the jet with independent orientation of magnetic fields and it can predict smaller variability for BL Lacs type of blazars 
(characterized by more than one emission regions). The emission regions in the jet can be filled with relativistic electron populations 
of different energy distributions \citep{Jones1985,Bjornsson1985}. Here, we consider only the depolarization effect of host galaxy contamination 
in R band to estimate the intrinsic jet polarization using the observed degree of polarization in the wavelength band 500-700 nm from the SPOL 
observations.

\subsection{Linear Synchrotron Polarization}
\begin{figure}
\epsscale{.75}
\plotone{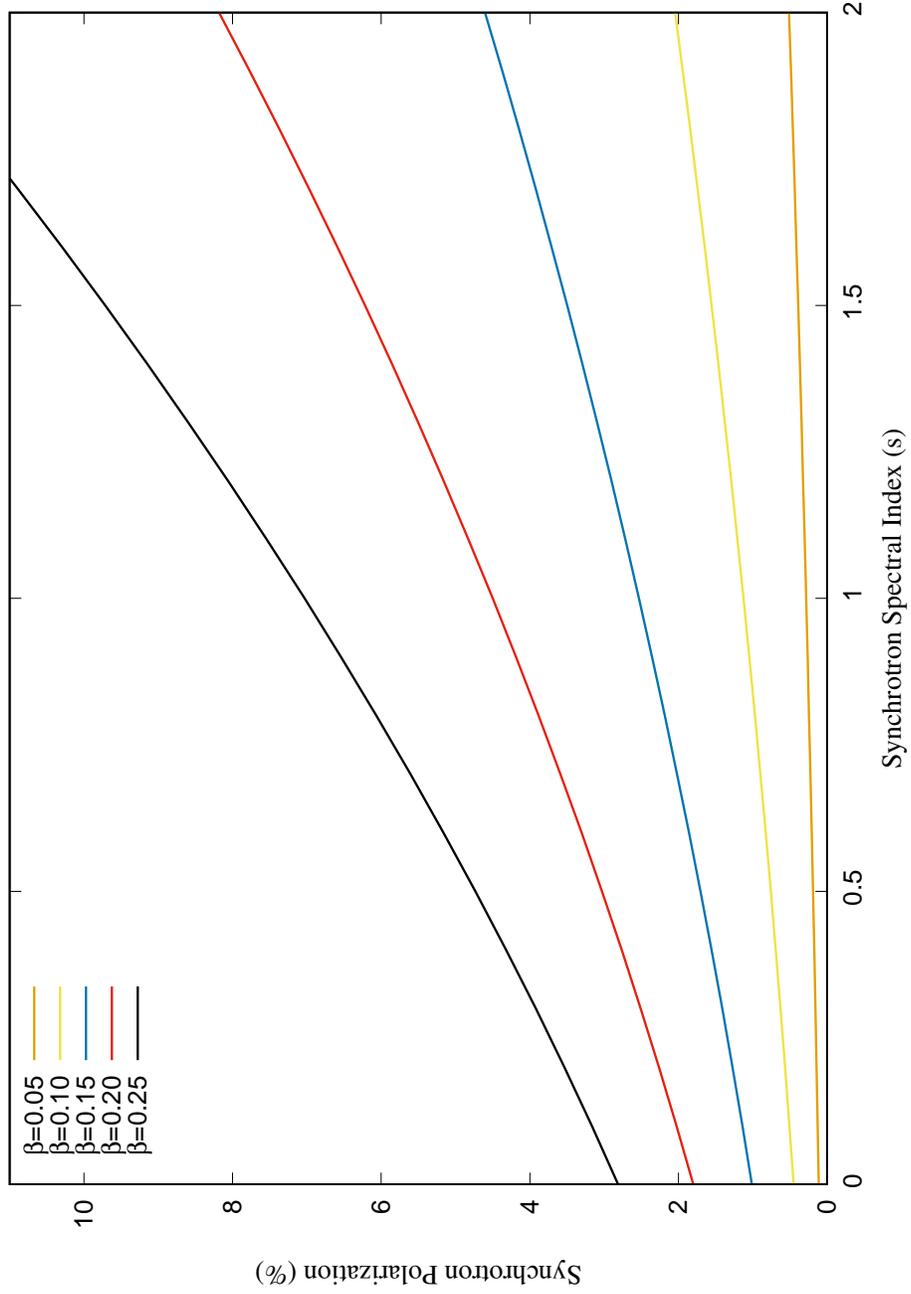}
\caption{Theoretical  degree of linear synchrotron polarization ($\Pi_{linear}$) as a function of synchrotron spectral 
	index for different configurations of the ordered and chaotic magnetic fields in the emission region.} 
\label{fig4}
\end{figure}

For blazars, it has been widely accepted that the non-thermal emission at radio and optical frequencies is the synchrotron radiation due to 
ultra-relativistic leptons (electrons/positrons) in the magnetic field of localized regions which move relativistically along the jet axis. 
Owing to the synchrotron origin, the radiation emitted in these frequency bands is expected to be highly polarized depending on the nature 
and geometry of the magnetic field in the emission region. The general formalism for estimating the degree of polarization of the radiation 
produced by the synchrotron process is properly understood in the literature \citep{Westfold1959,Legg1968}. The degree of polarization measured 
at radio wavelengths is observed to be less than that in optical bands. This suggests that the optical emission is produced from a smaller 
region with more uniform and ordered magnetic field in comparison to the radio emission \citep{Jorstad2013}. In the present study, we consider 
a simple scenario that there are two emission regions in the jet of Mrk 421 during the flaring episode in February 2010 where optical emission 
comes from one region and X-ray and $\gamma$--ray are produced from another region. The optical emission observed in V and R bands is attributed 
to the synchrotron radiation of the relativistic electrons in the emission region. The electrons are assumed to be accelerated to relativistic 
energies by the well known first order Fermi acceleration process before injecting into the emission region. The energy spectrum of relativistic 
electrons in the emission region can be approximated by a power law of the form
\begin{equation}\label{electron-eqn}
	N(\gamma)=K\gamma^{-\alpha}~~;~~~~~~~~~~~\gamma_{min}\le \gamma \le \gamma_{max}	
\end{equation}	
where $\alpha$ is the electron spectral index and $\gamma_{min}$ and $\gamma_{max}$ are the minimum and maximum electron Lorentz factors 
respectively. The magnetic field in the emission region ($B$) is considered to be 
\begin{equation}\label{magfield}
	B = B_o + B_c	
\end{equation}	
where $B_o$ is the strength of ordered magnetic field produced by shocks and $B_c$ is the  strength of chaotic magnetic field. 
The synchrotron photons produced by the above population of electrons (Equation \ref{electron-eqn}) gyrating in the resultant 
magnetic field $B$ (Equation \ref{magfield}) is described by a power law with energy spectral index 
\begin{equation}
	s = \frac{\alpha-1}{2}
\end{equation}
The degree of linear polarization ($\Pi_{linear}$) for optically thin synchrotron radiation depends on the photon spectral index 
(energy distribution of electrons) and structure of the magnetic field. It is expressed as \citep{Westfold1959,Bjornsson1985,Fraija2017}
\begin{equation}
	\Pi_{linear} = \frac{(s+1)(s+2)(s+3)}{8(s+5/3)} \beta^2
\end{equation}	
where $\beta$ is the ratio of ordered to chaotic magnetic fields strength in the emission region. The synchrotron radiation from blazars 
in low energy bands is observed to be optically thin. The expected percentage of synchrotron polarization ($\Pi_{linear}$) as a function 
of the observed synchrotron spectral index ($s$) for different configurations of $B_o$ and $B_c$ is shown in Figure \ref{fig4}. It is 
evident from Figure \ref{fig4} that the degree of linear synchrotron polarization for a given synchrotron spectral index ($s$) 
increases monotonically with increase in the ratio of two magnetic fields in the optical emission region. The increasing value of the 
ratio ($\beta$) implies that the ordered magnetic field dominates over the chaotic magnetic field. A completely isotropic chaotic 
magnetic field produces synchrotron radiation with zero linear polarization whereas perfectly ordered uniform magnetic field will result 
in the maximum degree of linear synchrotron polarization of $\Pi_{linear}^{max}$ $\sim$ 69--75$\%$ for relativistic electrons with 
typical spectral indices of $\alpha~\sim$ 2--3. The chaotic magnetic field in the emission region can be compressed by the moving shock 
waves. If the optical emission region is highly magnetized, a fast moving shock leads to the significant variation in the polarization 
leaving the flux level unchanged \citep{Zhang2016}. 
\par
We compare the theoretical degree of linear polarization ($\Pi_{linear}$) shown in Figure \ref{fig4} with the intrinsic jet polarization 
($\Pi_{jet}$) calculated using the observed polarization ($\Pi_{obs}$) from the SPOL observations (Section 5.1). The results from the 
comparison of the theoretical and measured degree of polarization during different epochs are summarized in Table \ref{table3}. It is 
observed that the measured optical polarization in the wavelength range 500-700 nm by the SPOL is broadly consistent with the 
theoretically expected synchrotron polarization for spectral index of $s = 0.6$ corresponding to $\alpha = 2.2$ with different values of the 
ratio of two magnetic fields ($\beta$). If the spectral distribution of the relativistic electrons is assumed to be described by a 
power law with index $\sim$ 2.2 in the optical emission region, the change in degree of polarization during different epochs can be 
attributed to the relative variation in the strength of ordered and chaotic magnetic fields. The lowest value of the linear 
polarization ($\sim$ 1.6$\%$) measured near simultaneous X-ray and $\gamma$--ray peak around February 16-17, 2010 (MJD 55243-55244) 
indicates a sudden increase in the strength of chaotic magnetic field component over the ordered magnetic field ($\beta \sim$0.17) 
in the emission region. This can happen due to the motion of turbulent plasma which makes the ordered magnetic field tangled keeping 
the resultant magnetic field constant in the emission region \citep{Begelman1998}. 
\par
Long term optical polarization study in R band using eight years of data suggested that the blazar Mrk 421 exhibits an average degree of 
linear polarization of $\sim$ 3.5$\%$ with preferred polarization angle of $\sim$ 169$^\circ$ during the period February 2008 to May 2016 
\citep{Fraija2017}. The variability analysis of the polarization in R band indicates that the average values of the degree and angle of 
polarization can be explained using two different emitting regions where the dominant optical emission originates from a region permeated 
by a stable magnetic field. Authors have also reported a connection between the polarimetric variations and spatial changes of the magnetic 
field in relativistic jet of Mrk 421. They have not considered the period of X-ray and $\gamma$--ray outburst of the source detected in 
February 2010.

\section{Conclusion}
In this paper, we have studied the behavior of linear polarization in the wavelength range 500-700 nm measured by the 
spectro-polarimeter at the Steward Observatory during the extreme flaring activity of the blazar Mrk 421 detected around 
February 16-17, 2010 (MJD 55243-55244). This outburst of the source was observed by many ground and space based 
multi-wavelength instruments world wide. The flaring activity was detected to be dominant in the X-ray and $\gamma$--ray energy 
bands. A summary of the nature of optical linear polarization obtained from the present study is given in the following points:
\begin{itemize}

	\item The optical light curves (V and R bands) indicate a relatively higher emission state of the blazar Mrk 421 in 
	      January 2010 prior to the giant X-ray and $\gamma$--ray flare observed around February 16-17, 2010. During the 
	      broadband flaring state, the optical emission (V and R bands) does not show any change in the source activity. 
	     The radio emission at 15 GHz is also found to be broadly consistent with the long term quiescent state of the 
	      source like its optical counterpart. The analysis of the optical emission using 100 year historical light curve 
              of Mrk 421 suggests a variability period of $\sim$ 17 years in R band \citep{Fraija2017}. This explains the 
	     the long-term quiescent state of the source in optical bands.
		
	\item Variability analysis of the optical (V and R bands) and radio light curves using different parameters indicates 
	      the absence of significant variability and the emission is consistent with the quiescent state of Mrk 421 derived 
	      from the long term observations. The variability parameters estimated for the period without flaring episode have 
              values higher than that during the outburst due to relatively high optical emission state in January 2010. 
		
	\item Polarization measurements from the blazar Mrk 421 suggest an average value of $\sim$ 4.2$\%$ for the degree of 
	      linear polarization in the wavelength range 500-700 nm for the period without flaring activity of the source. 
	      The optical linear polarization during February 16-17, 2010 outburst is found to be $\sim$ 1.6$\%$ which is less 
	      than the average value of $\sim$ 2.2$\%$ for the week of X-ray and $\gamma$--ray outburst. This indicates that 
	      the degree of optical linear polarization decreases during the flaring activity of the source in the X-ray and 
	      $\gamma$--ray energy bands. The angle of polarization has an average value of $\sim$ 137$^\circ$ during the entire 
	      period. 
		
      \item   Variability analysis of the degree of linear polarization based on different variability parameters suggests that 
	      the optical polarization is variable during the X-ray and $\gamma$--ray outburst of Mrk 421. Whereas, angle of polarization 
	      does not change significantly. A non-zero positive Pearson coefficient ($\sim$0.20) for the measured degree of optical 
	      linear polarization with optical (V and R bands) and  radio fluxes is observed. This indicates possibly the common origin 
	      via the synchrotron process in the blazar jet.

      \item  The non-thermal contribution from the host galaxy of Mrk 421 in the R band is observed to play a significant role 
	     in depolarizing the linear synchrotron polarization intrinsic to the jet. The degree of linear polarization intrinsic 
	     to the jet is found to be approximately 1.5 times higher than the measured value in the wavelength range 500-700 nm by the 
	     SPOL after correcting for the contribution from the host galaxy. The estimated intrinsic linear polarization is consistent 
             with the synchrotron polarization produced by the relativistic electrons described by a power law with spectral index of 
	     $\sim$2.2 in the region filled with ordered and chaotic magnetic fields. The change in linear polarization can be attributed 
	     to the ratio of the strength of ordered magnetic field to the chaotic component in the emission region. However, simultaneous 
	      measurements of the degree of linear polarization in different optical bands during the flaring episodes are very important 
	     to fully explore the physical processes for blazars and effect of wavelength dependent Faraday rotation on the polarized jet 
	     emission.	
     
      \item   Our study suggests the presence of two emission regions in the jet of Mrk 421 where optical emission is 
	      produced from one region and X-ray and $\gamma$--rays are produced from another region. The optical emission 
	      region can be filled with ordered and chaotic magnetic fields and steady population of relativistic electrons. 
	      The decrease in optical linear polarization during the X-ray and $\gamma$--ray outburst is caused due to sudden 
	     dominance of the chaotic magnetic field over the ordered magnetic field in the emission region. Whereas, the X-ray 
	     and $\gamma$--ray flaring activity possibly originates due to the sudden increase in the density of relativistic electrons 
	     in the second emission region \citep{Singh2017}.  
	
\end{itemize}

\acknowledgments
We thank the anonymous reviewer for his/her useful suggestions to improve the contents of the manuscript. 
This research has made use of data from the OVRO 40-m monitoring program (Richards, J. L. et al. 2011, ApJS, 194, 29) which is 
supported in part by NASA grants NNX08AW31G, NNX11A043G, and NNX14AQ89G and NSF grants AST-0808050 and AST-1109911. Data from 
the Steward Observatory spectropolarimetric monitoring project were used. This program is supported by Fermi Guest Investigator 
grants NNX08AW56G, NNX09AU10G, NNX12AO93G, and NNX15AU81G.


\clearpage
\clearpage
\clearpage
\clearpage
\clearpage
\begin{table}
\begin{center}
\caption{Variability Parameters estimated during the period of the X-ray and $\gamma$--ray outburst February 13-19, 2010 (MJD 55240-55246)}
\vspace{1.0cm}
\begin{tabular}{lclll}
\tableline\tableline
Observables  		&$F_{var}$  		&$A_{mp}$(\%) 	&$RVA$ \\
\tableline
Optical flux (V) 	&0.17$\pm$0.05 	     &9.10$\pm$1.96 	&0.05$\pm$0.01 \\
Optical flux (R)        &0.17$\pm$0.05	     &5.86$\pm$1.56     &0.06$\pm$0.01 \\
Radio (15 GHz)          &0.34$\pm$0.18	     &--		&0.06$\pm$0.01 \\
Linear Polarization	&0.34$\pm$0.09	     &67.63$\pm$8.78	&0.55$\pm$0.07 \\
Polarization Angle      &0.17$\pm$0.04	     &28.40$\pm$1.21	&0.10$\pm$0.01	\\
\tableline
\end{tabular}
\label{table1}
\end{center}
\end{table}

\begin{table}
\begin{center}
\caption{Variability Parameters calculated excluding the period of the X-ray and $\gamma$--ray outburst}
\vspace{1.0cm}
\begin{tabular}{lclll}
\tableline\tableline
Observables  		&$F_{var}$  		&$A_{mp}$(\%) 	&$RVA$ \\
\tableline
Optical flux (V) 	&0.16$\pm$0.04 	     &31.62$\pm$3.07 	&0.16$\pm$0.01 \\
Optical flux (R)        &0.15$\pm$0.04	     &28.76$\pm$2.94    &0.15$\pm$0.01 \\
Radio (15 GHz)          &0.07$\pm$0.01	     &8.91$\pm$5.4	&0.05$\pm$0.01 \\
Linear Polarization	&0.17$\pm$0.03	     &41.19$\pm$2.26	&0.23$\pm$0.04 \\
Polarization Angle      &0.16$\pm$0.03	     &33.23$\pm$1.62	&0.17$\pm$0.01	\\
\tableline
\end{tabular}
\label{table2}
\end{center}
\end{table}

\begin{table}
\begin{center}
\caption{Summary of the results derived from the comparison of the theoretical degree of linear polarization ($\Pi_{linear}$) 
	 for $s=0.6$ with the measured polarization in the wavelength range 500-700 nm by the SPOL}
\vspace{1.0cm}
\begin{tabular}{lclll}
\tableline\tableline
Epoch		 &Time Interval~(MJD)			    &$\Pi_{obs} (\%)$ 	&$\Pi_{jet} (\%)$	&$\beta$\\
\tableline
Excluding Flare  &January 1 -March 31, 2010 (55197--55286)   &4.2 	     		&6.4 			&0.28 \\
During Flare     &February 13-19, 2010 (55240--55246)	     &2.2    			&3.4		        &0.20 \\
Near Flare       &February 16-17, 2010 (55243--55244)	     &1.6			&2.4	                &0.17 \\
\tableline
\end{tabular}
\label{table3}
\end{center}
\end{table}

\end{document}